\documentclass[12pt,preprint]{aastex}

\begin{document}

\title{Constraining the Structure of GRB Jets Through the Afterglow Light Curves}

\author{Jonathan Granot\altaffilmark{1} and Pawan Kumar\altaffilmark{2}}

\altaffiltext{1}{Institute for Advanced Study, Olden Lane,
Princeton, NJ 08540; granot@ias.edu}
\altaffiltext{2}{Astronomy
Department, University of Texas, Austin, TX 78731;
pk@surya.as.utexas.edu}

\begin{abstract}

We investigate the effect that the structure of GRB jets has on
the afterglow light curves for observers located at different
viewing angles, $\theta_{\rm obs}$, from the jet symmetry axis.
The largest uncertainty in the jet dynamics is the degree of
lateral energy transfer. Thus, we use two simple models, that make
opposite and extreme assumptions for this point, and calculate the
light curves for an external density that is either homogeneous,
or decreases as the square of the distance from the source.
The Lorentz factor and kinetic energy per unit solid angle are
initially taken to be power laws of the angle from the jet axis.
We perform a qualitative comparison between the resulting light
curves and afterglow observations. This constrains the jet structure,
and poses problems for a `universal' jet model, where all GRB jets
are assumed to be intrinsically identical, and differ only by our
viewing angle, $\theta_{\rm obs}$.

\end{abstract}


\section{Introduction}
\label{sec:intro}

There are several different lines of evidence in favor of
collimated outflows, also referred to as jets, in gamma-ray bursts (GRBs).
Perhaps, the best evidence so far is the achromatic break seen in the light
curves of many (though not all) GRB afterglows. A collimated outflow also
helps to reduce the estimate for the total energy output in gamma-rays that
is inferred from the fluence in GRBs with known redshifts, which in some
cases approaches and in one case (GRB 990123) even exceeds the rest energy
of a solar mass star, {for a spherically symmetric emission}.
Such an energy output in gamma-rays is hard to
produce in any model that involves a stellar mass object. Furthermore,
a non-spherical flow can also manifest itself through linear polarization
(Sari 1999; Ghisellini \& Lazzati 1999) as was indeed observed for a few
afterglows (Covino et al. 1999; Wijers et al. 1999; Rol et al. 2000).

Most GRB jet models consider an outflow that is uniform within some finite,
well defined, opening angle around its symmetry axis, and where the Lorentz
factor, energy density etc. drop sharply beyond this opening angle
(Rhoads 1997, 1999; Panaitescu \& M\'esz\'aros 1999;
Sari, Piran \& Halpern 1999; Kumar \& Panaitescu 2000;
Moderski, Sikora \& Bulik 2000; Granot et al. 2001, 2002).
Such a uniform jet with a sharp, well defined, edge shall be referred to
as a `top hat' jet. The possibility that GRB jets can display an angular
structure, where the kinetic energy per unit solid angle, $\epsilon$, and the
Lorentz factor, $\Gamma$, of the GRB outflow vary smoothly as power laws
in the angle, $\theta$, from the jet axis, was proposed by
M\'esz\'aros, Rees \& Wijers (1998). We shall refer to such smoothly
varying relativistic outflows as `structured' jets (as opposed to a
`top hat' jet, that has no inner structure).

Recently, several different groups have analyzed afterglow observations within
the frame work of the `top hat' jet model, and have inferred a relatively
narrow distribution both for the total energy output in gamma-rays
(Frail et al. 2001) and in the initial kinetic energy of the relativistic
outflow (Panaitescu \& Kumar 2001; Piran et al. 2001). These results
may alternatively be interpreted as GRB jets having a universal structure,
which is intrinsically the same for all GRBs, and the observed
differences between different GRBs are a result of different viewing
angle, $\theta_{\rm obs}$, w.r.t. the jet symmetry axis
(Lipunov, Postnov \& Prokhorov 2001; Rossi, Lazzati \& Rees 2002;
Zhang \& M\'esz\'aros 2002). Whereas in the `top hat' jet interpretation,
the jet break time, $t_j$, depends mainly on the initial opening angle of
the jet, $\theta_j$ (and also has a smaller dependence on its energy per
unit solid angle and on the external density, e.g. Sari, Piran \&
Halpern 1999), in the universal `structured' jet interpretation, $t_j$
depends mainly on the viewing angle, $\theta_{\rm obs}$, and the light curve
is similar to that for a `top hat' jet with an opening angle
$\theta_j=\theta_{\rm obs}$  and the same value of the energy per unit solid
angle of the `structured' jet at $\theta=\theta_{\rm obs}$.

While the evolution of `top hat' jets and their light
curves have been widely investigated (Rhoads 1999; Panaitescu \&
M\'esz\'aros 1999; Kumar \& Panaitescu 2000; Moderski, Sikora \& Bulik 2000;
Granot et al. 2002), including numerical simulations of the jet dynamics
and calculation of the resulting light curves (Granot et al. 2001),
much less work has been done on `structured' jets. In an accompanying paper
(Kumar \& Granot 2002) we calculate the dynamics of `structured'
relativistic jets by solving relativistic fluid dynamics equations, and
demonstrate that a simple analytic or semi-analytic model can qualitatively
reproduce the results for the jet dynamics and for the light curves as long
as the Lorentz factor along the jet axis is of order 4 or larger.
In this paper we use
two simple models for the jet dynamics, which are designed to bracket the
true hydrodynamic evolution of the jet, to calculate the afterglow light
curves for a wide range of parameters for the structured jet and the external
density. A qualitative comparison of the light curves with afterglow
observations provides constraints on the jet structure and density
profile in its immediate vicinity.

In \S \ref{sec:model} we present the physical model. We begin with
the initial conditions (\S \ref{IC}), then describe our two simple dynamical
models (\S \ref{JD}) and finally outline the procedure for calculating the
light curves (\S \ref{LC}).
Our results are presented in \S \ref{sec:results}, and in \S \ref{sec:dis}
we discuss our main conclusions.

\section{The Physical model}
\label{sec:model}

The jet is assumed to posses axial symmetry, so that all the hydrodynamic
quantities at a given lab frame time, $t$, depend only on the angle, $\theta$,
from the jet symmetry axis. The radial profile of the outflow is ignored in
this simple treatment, and the shocked material is approximated by a thin
shell that is located at the same radius, $R$, of the shock.

The observed afterglow light curves depend on the angular structure of
the GRB jet, its hydrodynamic evolution, and the viewing angle,
$\theta_{\rm obs}$,
w.r.t. the jet symmetry axis. One may therefore use the shape of the
afterglow light curves to constrain the angular structure of the jet,
as well as to infer the viewing angle, $\theta_{\rm obs}$. For a `top hat'
jet of opening angle $\theta_0$, $\theta_{\rm obs}<\theta_0$ is required
in order to observe the prompt gamma-ray emission. In this case the
differences in the light curves between different  $\theta_{\rm obs}$
in the range $0\leq\theta_{\rm obs}<\theta_0$ are rather small
(Granot et al. 2001, 2002). However, for structured jets we expect large
differences in lightcurves for different $\theta_{obs}$.

\subsection{The Initial Conditions}
\label{IC}

The initial conditions are chosen at a lab frame time, $t_0$, for
which on the one hand, the internal shocks have ended, and on the
other hand, no significant deceleration due to the sweeping up of the
external medium has yet occurred (i.e. $R_{IS}(\theta)<R(\theta,t_0)
<R_{\rm dec}(\theta)$ for all relevant $\theta$). For simplicity,
the original ejecta is assumed to remain cold, even near the
deceleration radius, where it is expected to be heated by the passage
of the reverse shock. This approximation might introduce inaccuracies
of order unity near the deceleration epoch, {corresponding to an observed
time $\lesssim$} a few minutes for a typical GRB, but it should have no 
effect on lightcurves at later times. Energy conservation implies
\begin{equation}
\label{eneregy2}
\epsilon(\theta,t_0)=\mu_0(\theta,t_0)\left[\Gamma(\theta,t_0)-1\right]
+\mu_s(\theta,t_0)\left[\Gamma^2(\theta,t_0)-1\right]
\approx \mu_0(\theta,t_0)\Gamma(\theta,t_0)\ ,
\end{equation}
where $\mu_0$ and $\mu_s$ are the rest mass per unit solid angle of the
original ejecta and of the swept up external medium, respectively,
$\epsilon$ is the energy (not including the energy associated with the
rest mass) per unit solid angle in the outflow, $\Gamma$ is the bulk
Lorentz factor of the shocked material, and $t_0$ is chosen so that
$\mu_s(\theta,t_0)\Gamma(\theta,t_0)\ll\mu_0(\theta,t_0)$ for all $\theta$.

The initial kinetic energy per unit solid angle, and initial Lorentz factor
(minus 1) are assumed to be power laws in $\theta$, outside of a core of
opening angle $\theta_c$:
\begin{equation}
\label{initial_con1}
\epsilon(\theta,t_0)=\epsilon_0\Theta^{-a} \quad,\quad
\Gamma(\theta,t_0)=1+(\Gamma_0-1)\Theta^{-b}\ ,
\end{equation}
where $\epsilon_0$ and $\Gamma_0$ are the initial kinetic energy per unit
solid angle and Lorentz factor at the jet axis, and
\begin{equation}\label{Theta}
\Theta\equiv\sqrt{1+\left({\theta\over\theta_c}\right)^2}\approx
\left\{\begin{array}{ll}1 & \theta\ll\theta_c \\
\theta/\theta_c & \theta\gg\theta_c \end{array} \right.\ .
\end{equation}
The velocity is assumed to be in the radial direction, so that
the initial radius is given by
\begin{equation}\label{R(t_0)}
R(\theta,t_0) = t_0\sqrt{1-\Gamma^{-2}(\theta,t_0)}\ ,
\end{equation}
and the lateral transfer of matter can be neglected\footnote{A more
rigorous treatment of the jet hydrodynamics (Kumar \& Granot 2002)
shows that these assumptions are reasonable.}.
The external mass density profile is assumed to be a power law in
the distance $r$ from the source, $\rho_{ext}(r)=Ar^{-k}$.
This implies
\begin{eqnarray}
\label{mu_s}
\mu_s(\theta,t)&=&\int_{0}^{R(\theta,t)}r^2dr\rho_{ext}(r)
={AR(\theta,t)^{3-k}\over(3-k)}\ ,
\\
\label{mu_0}
\mu_0(\theta,t)&=&\mu_0(\theta,t_0)=
{\epsilon_0\,\Theta^{b-a}\over(\Gamma_0-1)}
-\left[\Gamma(\theta,t_0)+1\right]\mu_s(\theta,t_0)\ .
\end{eqnarray}

\subsection{The Jet Dynamics}
\label{JD}

The main uncertainty in the jet dynamics is the degree of lateral
transport of energy, from small to large angles, $\theta$, w.r.t.
the jet axis. We therefore make two alternative and extreme assumptions
regarding this transport: in {\bf model 1} we assume that the energy per unit
solid angle, $\epsilon$, does not change with time,
$\epsilon(\theta,t)=\epsilon(\theta,t_0)$, while in {\bf model 2} we
assume the maximal averaging of $\epsilon$ over the angle $\theta$ that
is consistent with causality. This is done by averaging over the
initial distribution of $\epsilon$,
\begin{equation}\label{model2_epsbar}
\bar\epsilon(\theta,t)\equiv{1\over(\cos\theta_{-}-\cos\theta_{+})}
\int_{\theta_{-}}^{\theta_{+}} d\theta'\sin\theta'\,\epsilon(\theta',t_0)\ ,
\end{equation}
where $\theta_{-}$ ($\theta_{+}$) is the angle below (above) $\theta$
(i.e. $\theta_{-}<\theta<\theta_{+}$), to which a hypothetical sound wave
that originated at $\theta$ at the initial time $t_0$ would have
propagated to. Initially, $\theta_{-}(t_0)=\theta_{+}(t_0)=\theta$,
and these angles subsequently evolve according to
\begin{equation}\label{theta_pm}
{\partial\theta_\pm\over\partial t}=
\pm{c_s(\theta_\pm,t)\over\Gamma(\theta_\pm,t)R(\theta_\pm,t)}\quad,\quad
c_s=\sqrt{\hat{\gamma}(\hat{\gamma}-1)[\Gamma-1]\over
1+\hat{\gamma}[\Gamma-1]}\ ,
\end{equation}
where $c_s$ is the local sound speed, and
$\hat{\gamma}=(4\Gamma+1)/3\Gamma$ is the adiabatic index. This simple
local scheme does not, in general, conserve the total energy in the outflow.
The global energy conservation is imposed by renormalizing $\bar\epsilon$,
\begin{equation}\label{model2_eps}
\epsilon(\theta,t)=\bar\epsilon(\theta,t)
{\int_0^{\pi/2}d\theta'\sin\theta'\,\epsilon(\theta',t_0)
\over\int_0^{\pi/2}d\theta'\sin\theta'\,\bar\epsilon(\theta',t)}\ .
\end{equation}

For both models 1 and 2, the radius of the thin shell of matter changes
as\footnote{There is a small difference between the velocity of the shock
front and that of the fluid just behind the shock (Blandford \& McKee 1976),
however we neglect this distinction in the present treatment, for simplicity.}
\begin{equation}\label{dRdt}
{\partial R(\theta,t)\over\partial t}=\sqrt{1-\Gamma^{-2}(\theta,t)}\ ,
\end{equation}
while $\mu_s$ and $\mu_0$ are given by equation (\ref{mu_s}) and
(\ref{mu_0}), respectively, and the Lorentz factor of the shocked material,
$\Gamma$, is obtained by solving the equation for energy conservation,
\begin{equation}\label{Gamma}
\epsilon=(\Gamma-1)\mu_0 + (\Gamma^2-1)\mu_s\quad,\quad
\Gamma={\mu_0\over 2\mu_s}
\left[\sqrt{1+{4\mu_s(\epsilon+\mu_0+\mu_s)\over\mu_0^2}}-1\right]\ .
\end{equation}

\subsection{The Light Curves}
\label{LC}

The local emissivity is calculated using the conventional assumptions
of synchrotron emission from relativistic electrons that are accelerated
behind the shock into a power law distribution of energies,
$N(\gamma)\propto\gamma^{-p}$ for $\gamma>\gamma_m$, where the electrons
and the magnetic field hold fractions $\epsilon_e$ and $\epsilon_B$,
respectively, of the internal energy. The shape of the local spectral
emissivity is approximated as a broken power law with breaks at the
typical synchrotron frequency $\nu_m$ and at the cooling frequency $\nu_c$
(Sari, Piran \& Narayan 1998). As our main focus in this work is the effect
of the jet dynamics on the afterglow emission, we keep the expression for
the local emissivity fairly simple, and leave the inclusion of additional
features such as the self absorption frequency, inverse Compton emission
and its effect on the electron cooling, etc., for later applications,
since such complications might make it hard to pinpoint the effects of
the jet dynamics on the light curves.

The light curves for observers located at different angles,
$\theta_{\rm obs}$, w.r.t. the jet axis, are calculated by applying
the appropriate relativistic transformation of the radiation field from
the local rest frame of the emitting fluid to the observer frames, and
integrating over equal photon arrival time surfaces (Granot, Piran \& Sari
1999; Kumar \& Panaitescu 2000; Granot et al. 2001). The radiation calculation
is in this sense rigorous and takes into account all relevant effects,
so that the resulting light curves accurately reflect what is expected for
a given jet structure and dynamics, where the latter are much less certain.

\section{Results}
\label{sec:results}

\begin{figure}
\centering
\noindent
\includegraphics[width=14cm]{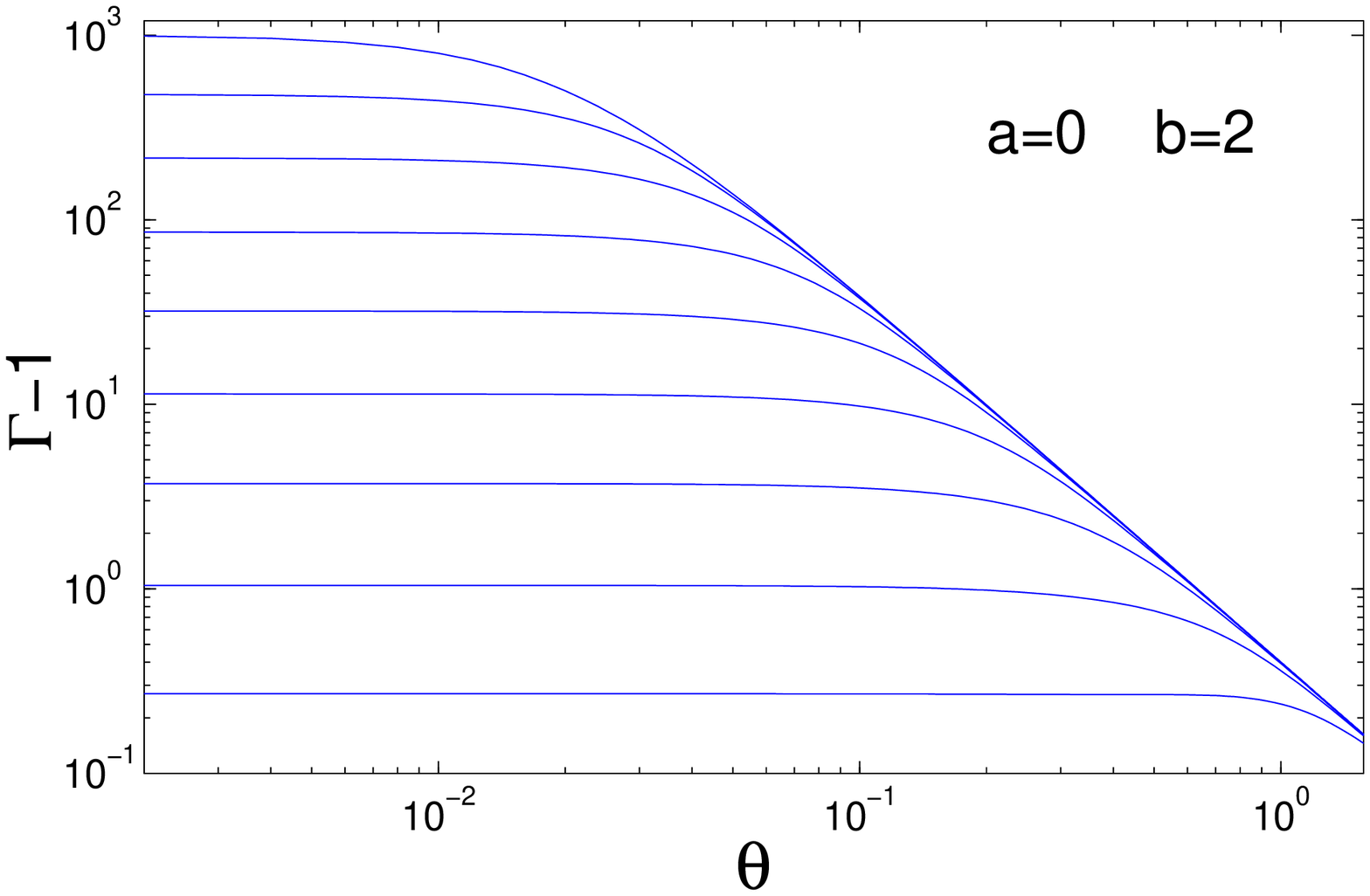}
\caption{\label{hydro_a0b2} The evolution of the Lorentz factor
$\Gamma(\theta)-1$, for $(a,b)=(0,2)$, $\Gamma_0=10^3$ and $\theta_c=0.02\;$rad.
For $a=0$, models 1 and 2 become the same,
so that this figure applies to both models.}
\end{figure}

\begin{figure}
\centering
\noindent
\includegraphics[width=14cm]{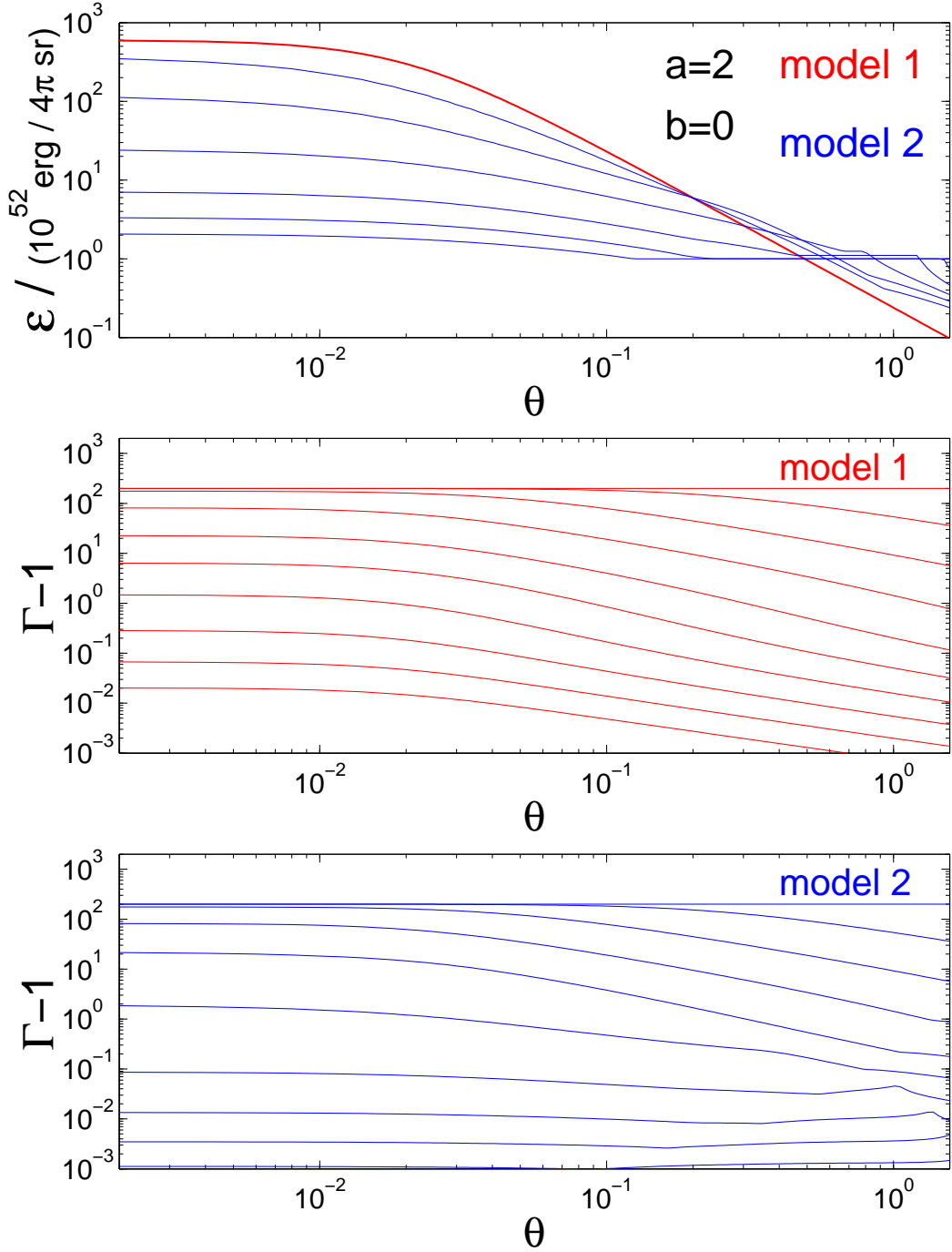}
\caption{\label{hydro_a2b0} The hydrodynamic evolution of the energy per
unit solid angle $\epsilon$ and the Lorentz factor $\Gamma-1$, as
a function of the angle $\theta$ from the jet symmetry axis,
according to our two simple and extreme models, for $(a,b)=(2,0)$.}
\end{figure}

\begin{figure}
\centering
\noindent
\includegraphics[width=14cm]{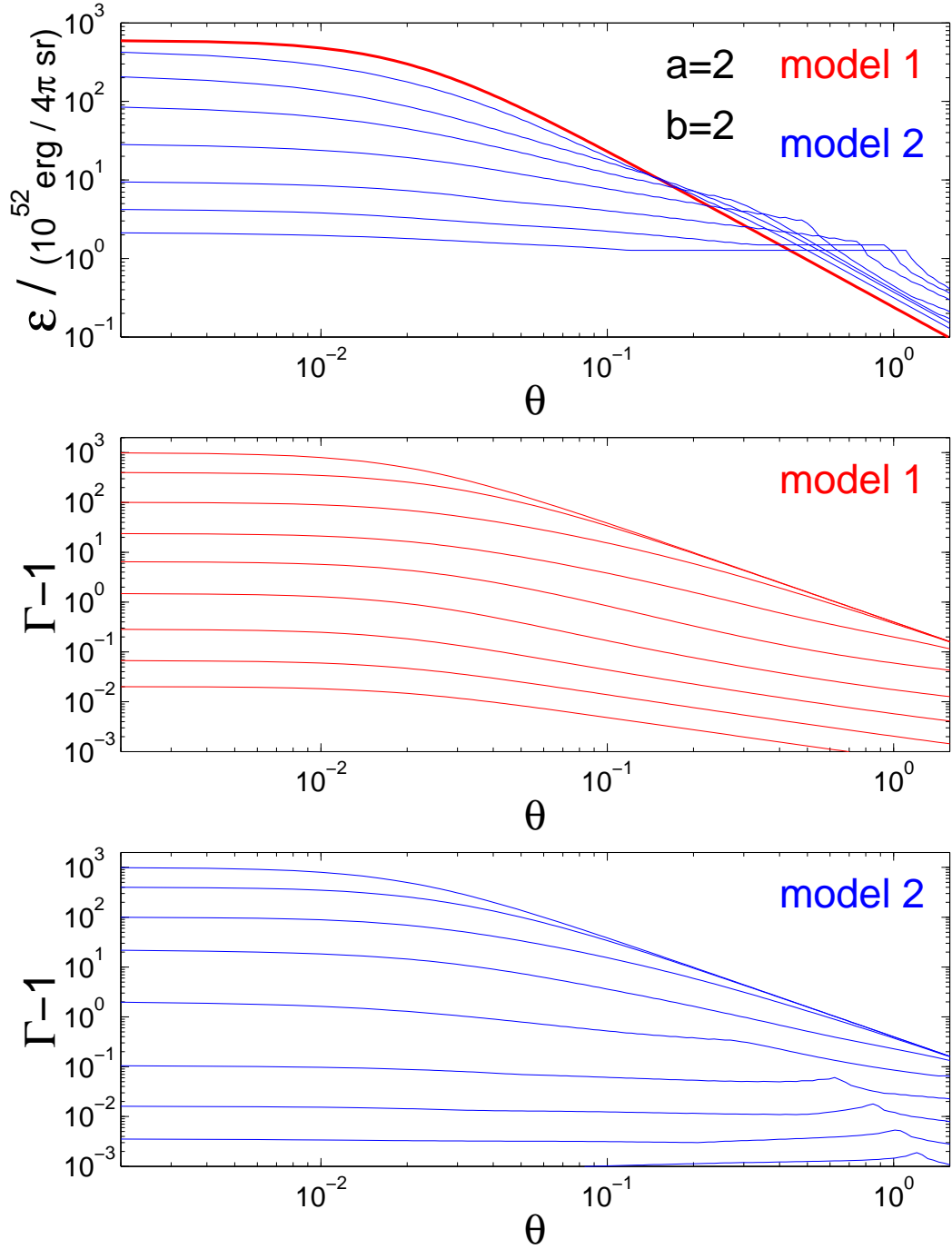}
\caption{\label{hydro_a2b2} The same as figure \ref{hydro_a2b2}
but for $(a,b)=(2,2)$.}
\end{figure}

In Figures \ref{hydro_a0b2}-\ref{hydro_a2b2} we show the evolution
of the hydrodynamic quantities according to the two models
described in the previous section, for a constant density external
medium ($k=0$). In Figure \ref{hydro_a0b2} we show the evolution
of the Lorentz factor for $(a,b)=(0,2)$. For $a=0$, models 1 and 2
become identical, so that this figure applies to both models.
Since $\epsilon(\theta,t)=\epsilon_0=\;$const, the evolution after
the deceleration time is the same for all $\theta$, and different
$\theta$ differ just by the deceleration time, $t_{\rm dec}$, and
the corresponding deceleration radius, $R_{\rm dec}$, that are
given by $\mu_s\sim\mu_0/\Gamma\sim\epsilon/\Gamma^2$, or
\begin{equation}\label{dec}
R_{\rm dec}=\left[{(3-k)\epsilon_0\over A c^2\Gamma_0^2}\right]^{1\over(3-k)}
\Theta^{2b-a\over(3-k)}\quad,\quad
t_{\rm dec}={R_{\rm dec}\over 2c\Gamma^2(t_0)}\propto
\Theta^{2b(4-k)-a\over(3-k)}\ .
\end{equation}
As can be seen from equation (\ref{dec}), for $a=0$, the deceleration occurs
first at small $\theta$, and gradually proceeds to larger angles. This can be
nicely seen in Figure \ref{hydro_a0b2}.

This situation is similar for $(a,b)=(2,2)$, but is reversed for
$(a,b)=(2,0)$, as can be seen in Figures \ref{hydro_a2b0} and
\ref{hydro_a2b2}, where in the latter case the deceleration occurs
first at large angles $\theta$ and then proceeds to smaller
angles. This is in agreement with equation (\ref{dec}). The upper
panel of Figures \ref{hydro_a2b0} and \ref{hydro_a2b2} shows the
energy per unit solid angle, $\epsilon$, which is initially the
same for the two models, and remains unchanged for model 1, while
for model 2 $\epsilon$ decreases at small angles $\theta$,
and increases at large angles. This indicates a lateral
transfer of energy from small to large angles. At late times, as
the flow becomes sub-relativistic, $\epsilon$ becomes almost
independent of angle $\theta$, and the flow approaches spherical
symmetry, as is indeed expected to occur physically, since the
spherically symmetric Sedov-Taylor self similar solution should be
asymptotically approached in the non-relativistic regime. The
middle and bottom panels show the Lorentz factor minus one,
$\Gamma-1$, for models 1 and 2, respectively. For model 1, it can
be seen that while the flow is still relativistic, The Lorentz
factor soon settles into a profile of $\Gamma-1\propto\theta^{-1}$
at $\theta>\theta_c$, instead of the initial
$\Gamma-1\propto\theta^{-2}$ or $\Gamma=\;$const. This occurs
since the fraction of the energy in the original ejecta soon
becomes negligible implying $\epsilon\approx\mu_s(\Gamma^2-1)$ so
that as long as the flow is relativistic, $\mu_s\propto
R^{3-k}\approx\;$const, since $R\approx ct$ for all $\theta$, and
$\Gamma-1\propto\epsilon^{1/2}\propto\theta^{-1}$. When the flow
becomes non-relativistic, $\Gamma^2-1\approx
2(\Gamma-1)\approx\beta^2$ so that
$\Gamma-1\propto\theta^{-4/(5-k)}$, which is still quite close to
$\theta^{-1}$ for $k$ between 0 and 2. For model 2, a similar
effect is seen at early times, when
$\epsilon(\theta,t)\approx\epsilon(\theta,t_0)$, but it becomes
much more homogeneous at later times, as $\epsilon(\theta)$
becomes more uniform.

The light curves of models 1 and 2 for
$(a,b)=(0,0),\,(0,2),\,(2,0),\,(2,2)$ and a uniform external
medium ($k=0$) are shown in Figure \ref{LCk0}, while the temporal
decay slope, $\alpha$ (defined by $F_\nu\propto t^\alpha$), for
the same light curves, is shown in Figure \ref{alpha_k0}. We have
added the spherical case, $(a,b)=(0,0)$, for comparison. For
$(a,b)=(0,2)$, the light curve initially rises, before the
deceleration time, $t_{\rm dec}$, which occurs at latter times for
larger viewing angles $\theta_{\rm obs}$, in accord with equation
(\ref{dec}), when keeping in mind that the emission is dominated
by $\theta\sim\theta_{\rm obs}$ at early times due to the
relativistic beaming of the radiation emitted from the jet. {After
a time $t$, deceleration has occurred at $\theta<\theta_{\rm dec}(t)$,
where $\theta_{\rm dec}$ is given by $t_{\rm dec}[\theta_{\rm dec}(t)]=t$,
and the light curves for $\theta_{\rm obs}<\theta_{\rm dec}(t)$ approach
the spherical case, and hence become very close to one other.
Keeping $a=0$, $\epsilon$ and $\Gamma_0$ fixed, and increasing $b$ from zero,
$t_{\rm dec}(\theta_{\rm obs})$ begins to shift to larger times,
and the ratio of the deceleration times for two given values of $\theta_{\rm obs}$
grows as $b$ increases. But after the time $t_{\rm dec}(\theta_{\rm obs})$,
the light curves still approach the same spherical light curve, for all values
of $b$. There is no jet break in the light curve for $a=0$.}
Since jet breaks are observed in many afterglows, this type of jet
structure can be ruled out as a universal model for GRB jets.

For $(a,b)=(2,0)$ and $(2,2)$ we find a jet break in the light
curve at roughly the same time as predicted in previous works
(Rossi, Lazzati \& Rees 2002; Zhang \& M\'esz\'aros 2002), i.e.
when $\Gamma(\theta=\theta_{\rm obs})\sim\theta_{\rm obs}^{-1}$.
However, we also find some new and interesting features.
For $(a,b)=(2,2)$ with $\Gamma_0\lesssim 10^3$,
the initial Lorentz factor at large
angles, $\theta$, is quite modest, resulting in a large
deceleration time, $t_{\rm dec}$, which can be as large as $t_{\rm
dec}\gtrsim 1\;$day for $\theta_{\rm obs}\gtrsim 0.2$. This would
result in a rising light curve at early times, for all
frequencies. On the other hand, for $(a,b)=(2,0)$, where the
initial Lorentz factor is the same for all $\theta$, the
deceleration time, $t_{\rm dec}$, is very small everywhere, so
that the initial rise of the light curve at $t<t_{\rm dec}$ will
be very hard to observe. The lack of an observation of a rising
light curve for afterglow observations starting from a few hours
after the burst can already constrain the jet structure: either
$b<2$ or $\Gamma_0>10^3$ are required. Future observations at much
earlier times after the burst, as may be achieved with the
forthcoming Swift mission, may provide much stronger constraints
on the jet structure.

It can also be seen from Figure \ref{LCk0}  that for $(a,b)=(2,2)$,
the value of the temporal decay slope, $\alpha$ (defined by $F_\nu\propto
t^{\alpha}$), before the jet break, is higher for larger viewing
angles $\theta_{\rm obs}$. This effect is very large for model 1,
but not very significant for model 2 (see Figure \ref{alpha_k0}).
The value of $\alpha$ before the break in the light curve for
model 1 is larger than the observed value for several well
studied GRB afterglows, and the light curve power-law
steepening, $|\delta\alpha|$, is smaller than the observed value
for many GRBs. The light curves obtained using model `1' at early
times are almost identical to the light curves from hydrodynamical
simulations, and this suggests that jet structure described by
(a,b)=(2,2) is not consistent with the observations for a number
of GRBs.

For $(a,b)=(2,0)$ the correlation between $\alpha$ before the jet
break and $\theta_{obs}$ is not seen, but instead there is a
flattening of the light curve (i.e. an increase in $\alpha$) just
before the jet break, for viewing angles sufficiently large
compared to the core angle of the jet, $\theta_{\rm obs}\gtrsim
3\theta_c$. This effect is more prominent in model 1, where it can
also be seen to a lesser extent for $(a,b)=(2,2)$ as well, compared 
to model 2, where this effect is smaller and can be seen only for 
$(a,b)=(2,0)$ (see Figure \ref{alpha_k0}). Light curves
calculated for the same jet profiles but with the jet dynamics
given by a hydro-simulation (Kumar \& Granot 2002) show that for
$(a,b)=(2,0)$ a reasonably sharp jet break is obtained only for
$\theta_{\rm obs}\gtrsim(2-3)\theta_c$, while the flattening in
the light curve just before the jet break becomes prominent at
$\theta_{\rm obs}\gtrsim(5-7)\theta_c$. This leaves only a factor
of $\sim 3$ in $\theta_{\rm obs}/\theta_c$ for which there is a
sharp jet break with no flattening in the light curve just before
this break, as is seen in all afterglows with jet breaks. However,
the inferred values of $\theta_{\rm obs}$ range at least one order of
magnitude, $\theta_{\rm obs}\sim 2-20^\circ$. 
Therefore, $(a,b)=(2,0)$ is unlikely to be a universal jet profile, 
for all GRBs, since $\theta_c$ is expected to vary between different 
afterglows. Even without requiring a universal jet profile, i.e.
the same $\theta_c$ for all GRB jets, it would be highly
improbable if all viewing angles would by chance fall within the
narrow range that gives a light curve qualitatively similar to
observations. This can exclude $(a,b)=(2,0)$ for most GRBs, and
imply $b>0$ (or even $b\gtrsim 1$). It is interesting to note that
this provides a lower limit on $b$, while the lack of detection of
a rising light curve at early times may provide an upper limit on
$b$. This may eventually pin down the value of $b$, or
alternatively, call into question the simple jet structure
interpretation discussed in this paper if the data is inconsistent
with a reasonable value for $b$.

Figure  \ref{LCk2} shows the light curves for an external density
that drops as $r^{-2}$ ($k=2$). This corresponds to a stellar wind
of a massive star progenitor. In the pulsar wind bubble model of
GRBs, the effective external density can have different profile
which may be approximated as a power law with an index $k$ that
ranges between $0$ and $1$ (K\"onigl \& Granot 2002). Therefore
$k=0$ and $k=2$, are the extreme values of $k$ that may be
expected, and an intermediate value of $k$, resulting in an
intermediate behavior of the light curves, is possible.

For $k=2$ we see a break in the light curve for a spherical flow
($a=b=0$), at $t\sim 10\;$yr, corresponding to the
non-relativistic transition time, $t_{\rm NR}$. This can also be
seen in Figure \ref{alpha_k2}, which shows the temporal decay
slope, $\alpha$, for the same light curves shown in Figure
\ref{LCk2}. Generally, $t_{\rm NR}\propto E^{1/(3-k)}$, so that
for $k=2$ it is linear in $E$, and a more moderate total energy
of, say, $10^{51}\;$erg would give $t_{\rm NR}\sim 1\;$yr. We
extend the observer time for which we show the light curves in
order to illustrate that in the non-relativistic regime, all light
curves become similar. One should keep in mind, that physically,
one may expect an $r^{-2}$ external density profile only up to
some finite radius (for a stellar wind, this correspond to the
radius of the wind termination shock) while at much larger radii
we expect a roughly constant density medium. This would result in
modifications to the light curves shown here, however, this is
typically expected to effect the light curves only in the
non-relativistic regime ($t\gtrsim t_{\rm NR}$).

The deceleration time, $t_{\rm dec}$, is typically very small for
$k=2$ since the density at small radii is very large, and the
total swept up mass is only linear in radius. The jet break is
much smoother and less sharp compared to a constant density environment
($k=0$), as was found for a `top hat' jet considered by Kumar \&
Panaitescu (2000). There is a relatively sharp break only for
$(a,b)=(2,0)$, and even then it is hardly sharp enough to
reproduce the jet breaks observed in GRB afterglows. This
suggests $k<2$, or rather $k\lesssim 1$. It is interesting to note
that this is consistent with the external density profile expected
in the pulsar wind bubble model (K\'onigl \& Granot 2002; Guetta
\& Granot 2002), while this is not consistent with the density
profile expected in the collapsar model, $k=2$.

\section{Discussion}
\label{sec:dis}

We have calculated light curves from structured relativistic jets
(jets whose Lorentz factor and energy per unit solid angle vary
smoothly with angle from the jet axis) using two simple models for
the jet dynamics, motivated by the hydrodynamical simulation of
Kumar and Granot (2002), and as two limiting cases of energy
redistribution in the lateral direction.

The first model considers the energy per unit solid angle,
$\epsilon(\theta,t)$, to be time independent, so that each segment
of the jet evolves independently of the other parts of the jet, as
if it was part of a spherical flow. This is roughly consistent
with the results from
hydrodynamical simulations of structured jets for which $\epsilon$
varies slowly with $\theta$, where we find the transverse velocity
in the comoving frame to be small compared to the sound speed
throughout much of the jet evolution, as long as the Lorentz
factor, $\Gamma$, is of order a few or larger along the jet axis
(Kumar \& Granot 2002). This model is expected to be a good
approximation for calculating the jet dynamics and the resulting
afterglow light curves for the first few days
after a GRB.

The second model considers the maximum possible re-distribution of
energy in the transverse direction so as to reduce the
lateral-gradient of $\epsilon$. The energy per unit solid angle,
$\epsilon(\theta,t)$, is taken to be proportional to the average
over the initial distribution of $\epsilon$, $\epsilon(\theta,t_0)$,
within the area out to which a sound wave could have traveled
from $\theta$ since the initial time, $t_0$. This model is
likely to be more accurate in describing the late time behavior of
the jet and the light curves, when the transverse velocity becomes
of order the sound speed.

We have calculated light-curves for a number of different initial
jet structures, which are taken to be power-law profiles at angles
larger than some core angle, i.e. $\epsilon\propto \theta^{-a}$
and $\Gamma\propto \theta^{-b}$ for $\theta>\theta_c$. We have
considered three different jet profiles, namely $(a,b)$ = (2,0),
(2,2), (0,2) for both a homogeneous external medium as well an
ambient density falling off as distance squared from the center of
explosion (i.e. $k=0$ or $2$, where $\rho_{ext}\propto r^{-k}$).

The quantitative differences in the light-curves calculated using
our two models are typically of order unity for all of the jet
structures we have considered. This gives us some confidence that
lightcurves can be calculated with a reasonable accuracy even with
a crude modeling of jet dynamics. There are, however, interesting
qualitative differences in the light-curve properties depending on
how we model jet dynamics. For instance, for $(a,b)=(2,2)$ and a
homogeneous medium ($k=0$), the temporal index of the light curve,
$\alpha$ (defined by $F_\nu\propto t^\alpha$), before the jet
break is larger for large viewing angles, $\theta_{\rm obs}$. This
effect is much more prominent for model 1, compared to model 2.
For $(a,b)=(2,0)$ and $k=0$, with model 1 there is a pronounced
flattening in the light curve just before the jet break for
$\theta_{\rm obs}\gtrsim 3\,\theta_c$, while for model 2 this
effect is much smaller.

\begin{figure}
\centering
\noindent
\includegraphics[width=14cm]{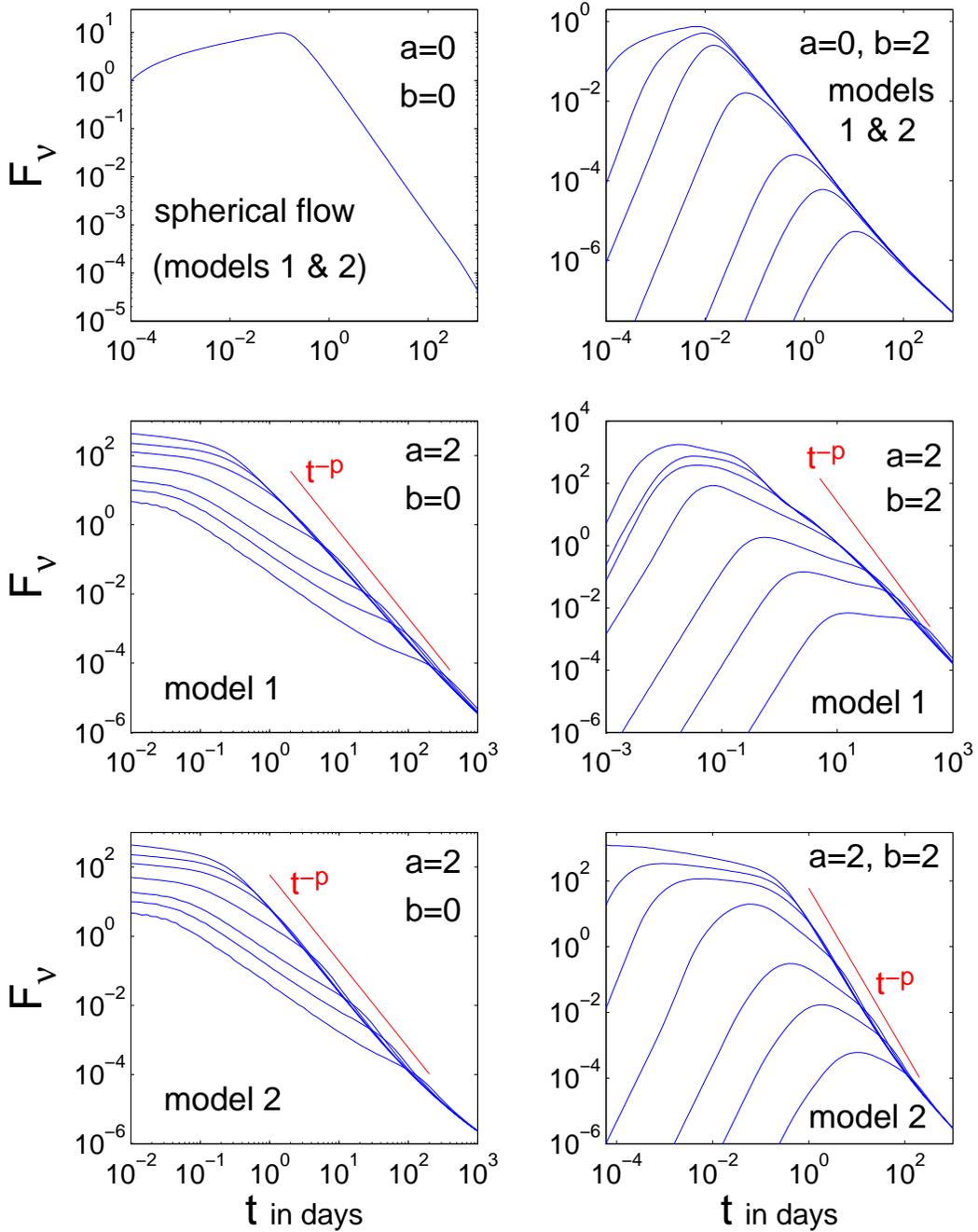}
\caption{\label{LCk0} Light curves for a constant density external medium
($k=0$), for models 1 and 2 (see text for details), in the optical
($\nu=5\times 10^{14}\;$Hz), for a jet core angle $\theta_c=0.02$,
viewing angles $\theta_{\rm obs}=0.01,0.03,0.05,0.1,0.2,0.3,0.5$,
$p=2.5$, $\epsilon_e=\epsilon_B=0.1$, $n=1\;{\rm cm}^{-3}$,
$\Gamma_0=10^3$, and $\epsilon_0$ was chosen so that the total energy of
the jet would be $10^{52}\;$erg. A power law of $t^{-p}$ is added in some
of the panels, for comparison.
}
\end{figure}

\begin{figure}
\centering \noindent
\includegraphics[width=14cm]{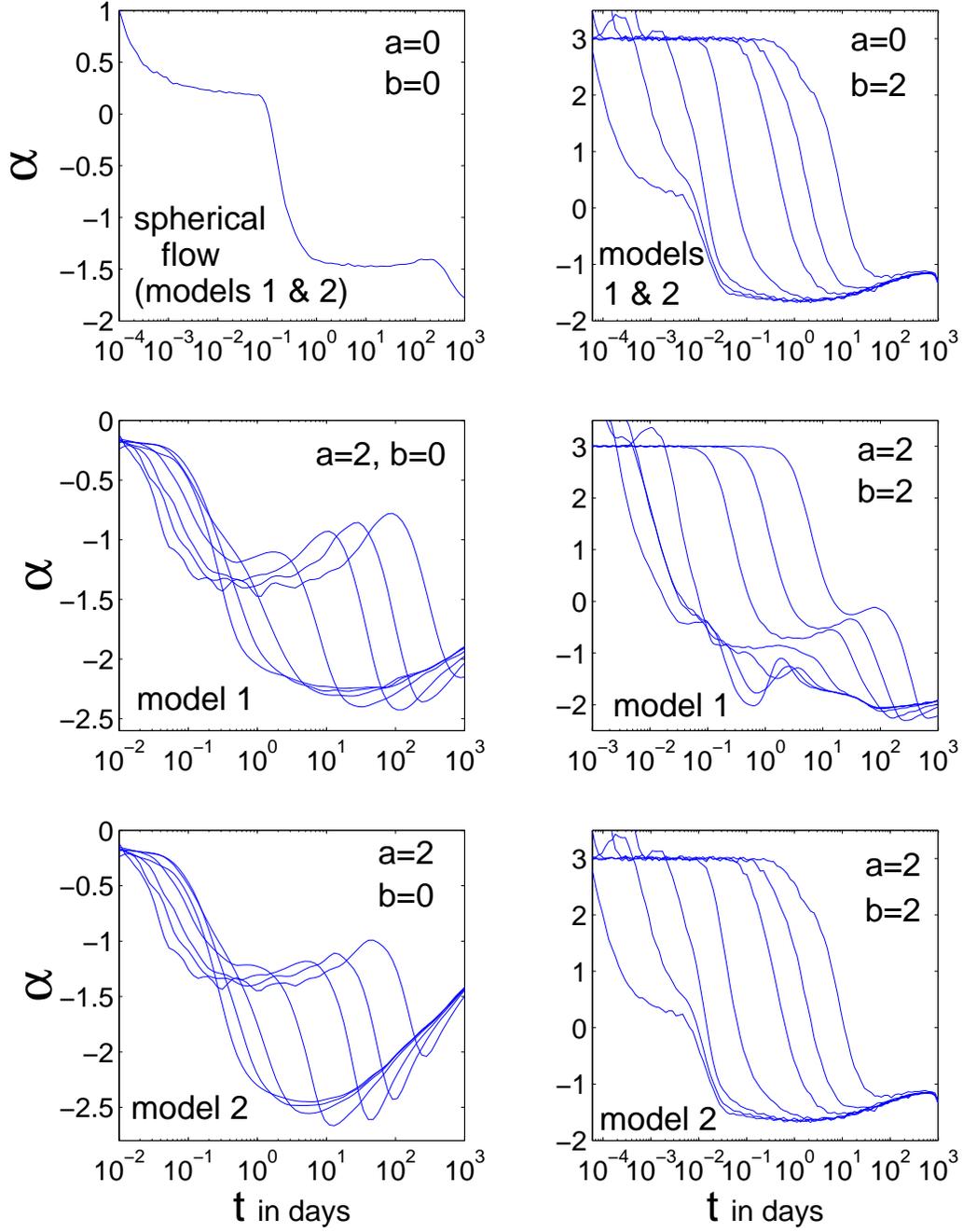}
\caption{\label{alpha_k0} The temporal decay slope, $\alpha\equiv
d\ln F_\nu/d\ln t$ (i.e. $F_\nu\propto t^\alpha$), for the light
curves shown in figure \ref{LCk0}, that are with a constant
density environment ($k=0$). }
\end{figure}

\begin{figure}
\centering
\noindent
\includegraphics[width=14cm]{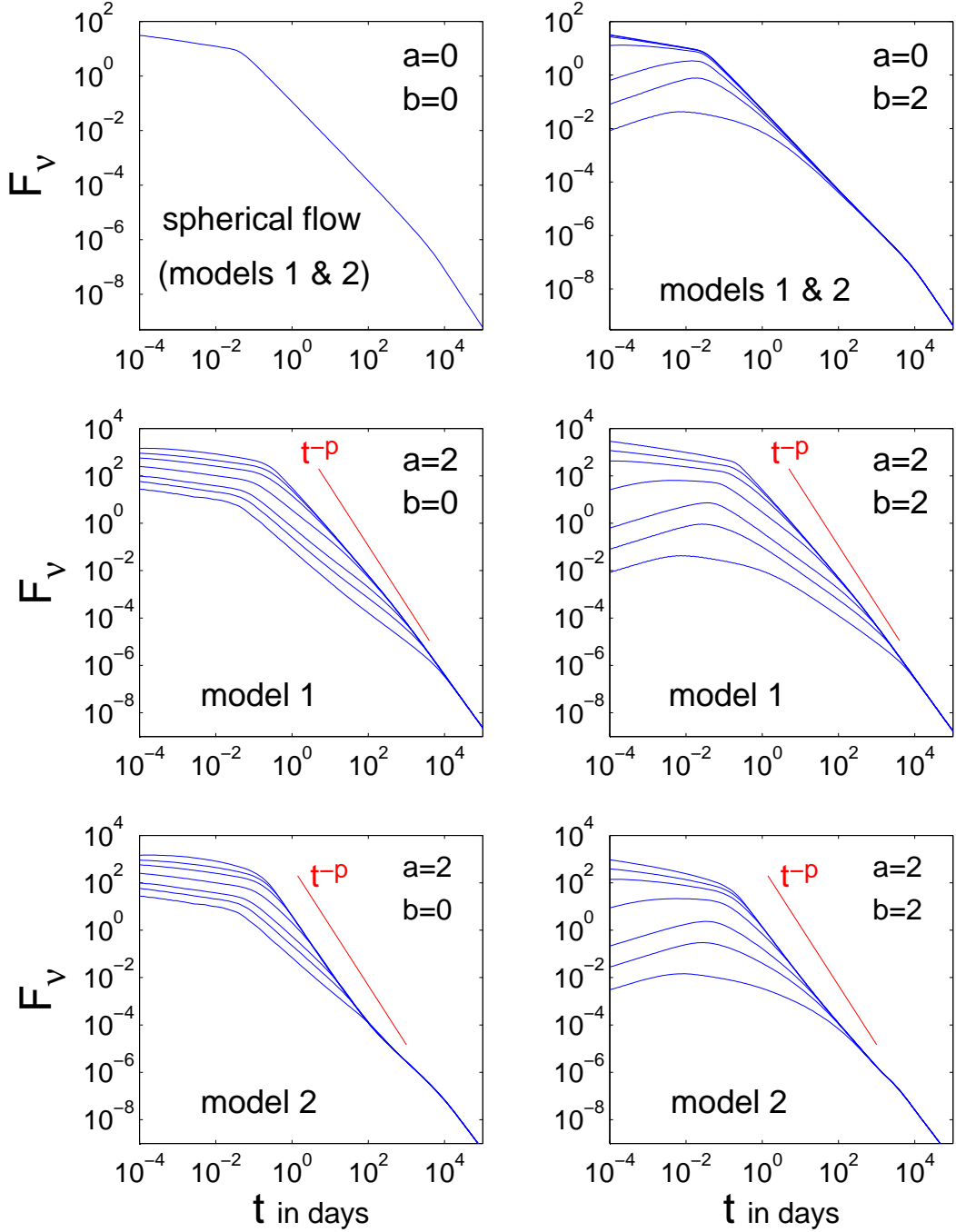}
\caption{\label{LCk2} The same as Figure \ref{LCk0}, but for an external density
$\propto r^{-2}$, corresponding to a stellar wind, with 
$A=5\times 10^{11}\;{\rm gr\; cm^{-1}}$.}
\end{figure}

\begin{figure}
\centering
\noindent
\includegraphics[width=14cm]{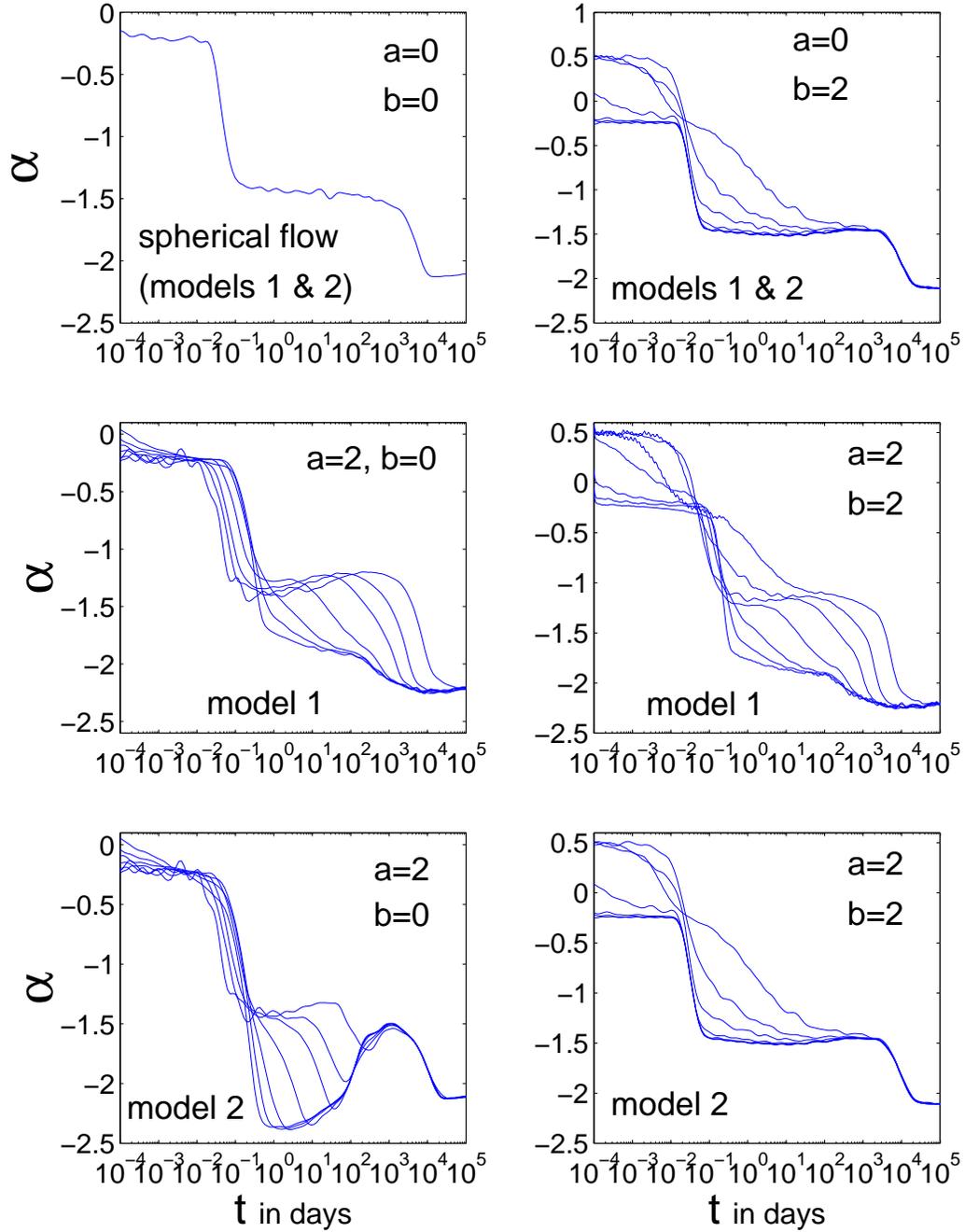}
\caption{\label{alpha_k2} The temporal decay slope, $\alpha$, for the light curves
shown in figure \ref{LCk2}, that are for a stellar wind environment ($k=2$).
}
\end{figure}

There is a significant difference between the two models in the
late time light curves, as we expect: model 1 light curves
continue a power-law decline at late times, whereas model 2 light
curves show a slight flattening at late times resulting from
energy redistribution -- see Figures \ref{LCk0} and \ref{alpha_k0}
for homogeneous external medium with jet structures $(a,b)=(2,2)$
\& (2,0), and Figures \ref{LCk2} and \ref{alpha_k2} for the
stratified external medium the particular case of $(a,b)=(2,0)$.
Figures \ref{LCk2} and \ref{alpha_k2} also show that the break in
the light curves for a stratified external medium is very gentle
and the change to the power-law index $\alpha$ is small so long as
the observer is not too far the axis: $\theta_{\rm
obs}/\theta_c\lesssim 10$. This result is consistent with the work
of Kumar \& Panaitescu (2000), based on the analysis of a `top
hat' jet model, that it is very difficult to see a break in light
curves for jets in an $r^{-2}$ medium.

We can learn about the jet structure in GRBs from comparing our
theoretical light curves to GRB afterglow observations. For
instance, the light curves for $(a,b)=(2,0)$ and (2,2) behave very
differently prior to the jet break. This can be used to
discriminate between these possibilities (Figures \ref{LCk0} \&
\ref{alpha_k0}). For obvious reasons, the light curves for the
(0,2) case are very distinct and similar to a spherically symmetric
explosion, and do not show a jet break. This
jet structure can therefore be ruled out for all GRBs that show a
jet break in their afterglow light curves. Clear breaks in the
light curves can be seen for a stratified external medium density
only when the observer is located far away from the jet axis
($\theta_{\rm obs}\gtrsim 10\,\theta_c$), however, in this case
the GRB will be faint and is likely to be missed in flux limited
triggers for detecting GRBs.

The light curves we obtain for structured jets, in many cases show
a different qualitative behavior compared to afterglow
observations. For example, $k=2$ does not produce a sufficiently
sharp jet break, while for $k=0$ only $a\approx 2$ produces jet
breaks that resemble afterglow observations, and even then, it is
very difficult to produce all current afterglow observations with
a universal jet profile, that is, if all GRB jets have the same
structure and differ only by our viewing angle, $\theta_{\rm
obs}$, w.r.t. the jet symmetry axis. Even if GRB jets do not have
a universal structure, a comparison between the observed light
curve and theoretical calculations can constrain the jet structure
of each GRB separately.

Future afterglow observations should enable us to determine the
structure of relativistic jets in GRBs as well as the properties
of the surrounding medium. This will enable us to determine
if the energy release in GRBs is nearly constant or not, and
whether the observed afterglow light curves and the jet break time
are determined by the observer angle w.r.t. a structured jet, or
by the opening angle of a `top hat' jet.

\acknowledgements

This work was supported by the Institute for Advanced Study, funds
for natural sciences (JG).

\newpage

\end{document}